\documentclass[letterpaper, 10 pt, conference]{ieeeconf}  

\IEEEoverridecommandlockouts                              
\usepackage{graphicx}      
\usepackage{subfigure}
\usepackage{enumerate}
\usepackage{amsmath}
\usepackage{amssymb}
\usepackage{booktabs}

\newtheorem{remark}{Remark}
\newtheorem{assumption}{Assumption}

\newtheorem{thm}{Theorem}

\title{\LARGE \bf Scenario optimization for optimal training of Echo State Networks}

\author{Luca Bugliari Armenio$^{1}$, Lorenzo Fagiano $^{1}$, Enrico Terzi$^{1}$, Marcello Farina$^{1}$, and Riccardo Scattolini$^{1}$
	\thanks{$^{1}$ The authors are with the Dipartimento di Elettronica, Informazione e Bioingegneria, Politecnico di Milano, Via Ponzio 34/5, 20133, Milano, Italy. E-mail: {\tt\small name.surname@polimi.it}}}

\begin{document}

\maketitle
\thispagestyle{empty}
\pagestyle{empty}

\begin{abstract}                
Echo State Networks (ESNs) are widely-used Recurrent Neural Networks. They are dynamical systems including, in state-space form, a nonlinear state equation and a linear output transformation. The common procedure to train ESNs is to randomly select the parameters of the state equation, and then to estimate those of the output equation via a standard least squares problem. Such a procedure is repeated for different instances of the random parameters characterizing the state equation, until satisfactory results are achieved. However, this trial-and-error procedure is not systematic and does not provide any guarantee about the optimality of the identification results. To solve this problem, we propose to complement the identification procedure of ESNs by applying results in scenario optimization. The resulting training procedure is theoretically sound and allows one to link precisely the number of identification instances to a guaranteed optimality bound on relevant performance indexes, such as the Root Mean Square error and the FIT index of the estimated model evaluated over a validation data-set. The proposed procedure is finally applied to the simulated model of a \emph{pH} neutralization process: the obtained results confirm the validity of the approach.
\end{abstract}


\section{Introduction}
In recent years the control community is experiencing a renewed interest in data-driven identification/learning (\cite{tan2018application}) and control (\cite{deisenroth2011learning}) techniques, fostered by the advent of more and more powerful machine learning tools and algorithms (\cite{pham1995neural}), (\cite{lee2000identification}), (\cite{bristow2006survey}). Indeed, the large availability of datasets enables effective modeling of complex dynamical systems (\cite{kutz2013data}).
Among the most employed tools, Neural Networks (NN) receive a particular attention for their flexibility in accomplishing identification tasks, spanning different classes of systems and problems (\cite{sutskever2014sequence}), (\cite{haykin2009neural}), (\cite{hinton2012deep}).

Several NN architectures have been proposed, among which Recurrent Neural Networks (RNN) are very promising due to their inherent capability to represent nonlinear dynamical systems.
Different methods to train RNN are present in the literature (\cite{jaeger2002tutorial}), all sharing the same core idea of minimizing the prediction error by tuning the available degrees of freedom of the network, i.e. its internal weights. However, it is not rare that the training algorithm results in a tough nonlinear problem, to be solved in an iterative (and time-consuming) way (\cite{pascanu2013difficulty}). In this paper, we focus on a particular class of RNN, i.e. the Echo State Networks (ESN) (\cite{jaeger2002tutorial}), that has been  successfully applied in many fields, such as speech recognition (\cite{pearlmutter1989learning}), time-series prediction (\cite{jaeger2004harnessing}), reinforcement learning (\cite{szita2006reinforcement}), and language modeling (\cite{tong2007learning}). In addition, ESN have recently been used for the design of Model Predictive Controllers with guaranteed stability properties (\cite{8728229}).

From a system theoretical standpoint, the structure of an ESN is composed of (i) a nonlinear state equation, where a sigmoid function is applied to a linear combination of states, inputs, and outputs, and (ii) an output transformation where the current output is a linear function of the states. The tuning of the ESN parameters usually consists of two phases. First, the parameters of the state equation are randomly generated; secondly, the output transformation parameters are computed solving a standard Least Squares (LS) problem. This procedure yields a dramatic reduction of the computational cost, however, the random choice of the state equation parameters naturally influences the final performance of the ESN, so that it is a common practice to repeatedly run the estimation procedure until a satisfactory result is achieved (\cite{jaeger2001echo}). As a consequence, no guarantees on the performance of the obtained ESN can be given, and suitable criteria to assess the quality of the trained network with respect to ``ideal'' results are not available.

In this paper, we solve this problem resorting to scenario optimization. Our contribution stems from a rather simple observation:  the described tuning procedure for ESN can be seen as a scalar optimization problem, where one wants to find the best performance index (no matter how defined, as long as it is systematically derived for a given network) among an infinite number of possible networks, each one featuring a random part - the state equation parameters - that is sampled according to a known probability distribution. This problem can not be solved exactly, however it can be turned into a convex, finite-dimensional optimization problem by means of a sampling approach, where the chosen network is picked as the best one among a finite number $N_\delta$ of trained ones. This falls perfectly into the framework of  scenario optimization theory (\cite{CAMPI2009149}), (\cite{Calafiore2005}), (\cite{campi2018introduction}), which provides a powerful result to certify the optimality of the chosen network with respect to a new one, obtained by sampling again from the same distribution and carrying out the LS identification phase. In particular, the probability that a newly trained network scores a better performance than the chosen one can be exactly computed on the basis of $N_\delta$. 
Based on this observation, in this paper we add the state equation parameters in the set of unknowns to be optimized, and solve the corresponding optimization problem in a random fashion, but with sound probabilistic optimality guarantees.\\
The proposed approach is finally tested on the problem of estimating the ESN model of a $pH$ neutralization process (\cite{hall1989modelling}), which represents a well recognized SISO non-linear  benchmark. The computed optimality guarantees are also verified \textit{a posteriori} over a large number of experiments, confirming the validity of the theoretical results.\\
%
\textbf{Notation}. Given a matrix A, we denote $A_{(i)}$ its $i^{th}$ row. $\|v\|$ is the 2-norm of vector $v$, $\otimes$ denotes the Kronecker product, $\mathbf{1}_{a,b}$ represents the matrix full of ones of dimensions $a,b$.

\section{Preliminaries on Echo State Networks}\label{Preliminaries}
The identification problem considered in this work consists in estimating a dynamical model that reproduces the behavior of an un unknown discrete-time plant, using $K$ measured samples of its inputs $u_{sys} \in \mathbb{R}^{n_u}$ and outputs $y_{sys} \in \mathbb{R}^{n_y}$. The considered model structure is that of ESN:
\begin{subequations}
	\begin{align}
	x(k+1) 			& = \zeta(W_xx(k)+W_{u}u(k)+W_yy(k))
	\label{net_state}\\
	\phi(k+1)		& =u(k)\\
	y(k) 			& = W_{out_{1}}x(k)+W_{out_{2}}\phi(k)
	\label{net_output}
	\end{align}
	\label{eq:RNN}\end{subequations}
where $k$ is the discrete time index, $x\in \mathbb{R}^{n}$ and $\phi \in \mathbb{R}^{n_u}$ are the model states, $u$ and $y$ are the inputs and outputs, respectively, $W_x, W_u, W_y, W_{out_{1}}$ and $W_{out_{2}}$ are weight matrices of proper dimensions, and $\zeta$ is a generic sigmoid Lipschitz continuous function, typically chosen as $\tanh(\cdot)$ (\cite{sohrab2003basic}).

The properties of the ESN \eqref{eq:RNN} have been studied in \cite{8728229}, where it has been proven that, if $\|W_x\|<1$, the system is Incrementally Input-to-State Stable ($\delta ISS$) with respect to inputs $u$ and $y$ (\cite{Discrete-timedeltaISS}). This guarantees that, running the system with the same inputs and different initial states, the transients associated to the initial conditions asymptotically vanish, thus enabling a consistent estimation procedure of the unknown parameter matrices $W_{out_{1}}$ and $W_{out_{2}}$. As discussed in \cite{kim2005note}, this property is strictly related to the one of fading memory. In addition, the $\delta ISS$ property is fundamental to design state observers and predictive regulators with stability guarantees, see again \cite{8728229}.\\
\\
\emph{Training procedure for ESN}\\
\\
Inspired by~\cite{jaeger2002tutorial}, the standard training of the ESN model \eqref{eq:RNN} proceeds according to the following steps:
\begin{enumerate}[(a)]
	\item collect from the plant the, possibly normalized (see \cite{jaeger2001echo}),  input and output sequences $u_{sys}$ and $y_{sys}$;
	\item define the system order $n+n_u$;
	\item generate a sparse matrix $W_x$, with elements sampled from a uniform distribution and such that $\|W_x\|<1$;\label{Step 3}
	\item generate random matrices $W_u$ and $W_y$ of proper dimensions,  with elements sampled from a uniform distribution;
	\item start from an arbitrary initial state $x(0)$ and run the state equation \eqref{net_state} forced by $u_{sys}$, $y_{sys}$; collect the computed state values $x(k)$;
	\item discard the first $K_0$ points of $u_{sys}$, $y_{sys}$, $x$ to remove the effects of the initial state (recall the $\delta ISS$ property guaranteed by the choice of $W_x$ in step \ref{Step 3});
	\item store the values of $\left(x(k),u_{sys}(k-1)\right)$ and $y_{sys}(k)$ for $k\geq K_0$ into matrices $\Phi$ and $Y_{sys}$ representing the output transformation \eqref{net_output} in vector form;
	\item solve the Least Squares problem
	\begin{equation*}
		\min\limits_{W_{out_{(i)}}} \|Y_{sys,i}-\Phi W_{out_{(i)}}'\|^2,
	\end{equation*}
	where $W_{out}=\left[W_{out_1} \quad W_{out_2}\right]$ and $Y_{sys,i}$ is the vector collecting all the samples pertaining to the $i^{th}$ output of $Y_{sys}$.\label{Step 8}
\end{enumerate}
Letting $\bar{K}=K-K_0$ and $Y=[y(K_0),y(K_0+1),\dots,y(K)]'$, the quality of the estimated model can then be evaluated with the root mean-square error (RMSE):
\begin{equation}
\label{RMSE}
RMSE = \sqrt{\frac{1}{\bar{K}}\|Y_{sys}-Y\|^2}
\end{equation}
Alternatively, a common index of the quality of the model is the FIT value:
\begin{equation}\small
\label{FIT}
FIT=100\cdot\left(1-\frac{\|Y_{sys}-Y\|}{\|Y_{sys}-\mathbf{1}_{\bar{K},1} \otimes \bar{y}\|}\right)\in (-\infty, 100]
\end{equation}
\normalsize
where $\bar{y}$ is the mean of the output of the system. As customary in identification (\cite{ljung1987system}), the error indexes \eqref{RMSE} and \eqref{FIT} should be computed over a validation dataset to actually assess the model performances. Given the definitions \eqref{RMSE} and \eqref{FIT}, the larger the $FIT$ or the smaller the $RMSE$, the better is the identified model quality.\\ 
Note however that, due to the adopted training procedure, the computed value $W_{out}$ of the output transformation parameters is itself a random variable, since it is a function of the random variables $W_x$, $W_u$, $W_y$. A sensible question is then: how many times shall one repeat the training procedure before being confident about the performance achieved by the best performing model among the generated ones? We answer to this question by providing a simple and effective guideline, based on 
the scenario optimization theory.
\\

	\section{The scenario approach}
The scenario approach is developed in the context of optimization in the presence of uncertainty. It considers the case of a convex program subject to constraints that depend, possibly in a nonlinear and non-convex way, on an uncertain variable $\delta \in \Delta$ characterized by a probabilistic, and possibly unknown, distribution. Specifically, consider the problem of maximizing a scalar function $f(\delta)$. In our specific application, such a function can be the FIT value obtained by the trained model.
We can write the optimization problem as follows.
\begin{equation}
\max_{\delta \in \Delta} f(\delta)
\label{eq:scenario_epigraphic}
\end{equation}
%
Problem \eqref{eq:scenario_epigraphic} is computationally intensive and possibly intractable due to the infinite number of constraints. For this reason, the scenario approach represents a viable solution. In this approach, a set of cardinality $N_{\delta}$ of uncertainty samples $\delta_i \in \Delta$, named \textit{scenarios}, are collected, and the following finite-dimensional problem is stated.
\begin{align}
h^*=\arg\min\limits_{h}&{\quad h} \notag\\
\text{s.t.}& \quad f(\delta_i)\leq h, \quad \forall i=1,\dots,N_{\delta}
\label{eq:scenario_samples}
\end{align}
Note that problem \eqref{eq:scenario_samples} is convex, and it can be rewritten as 
\[
h^*=\max\limits_{i=1,\dots,N_{\delta}}f(\delta_i).\]
Since the problem considers only a finite number of constraints, it is not guaranteed that the obtained solution is the absolute best one, however its optimality can be quantified in probabilistic terms, making the level of reliability of the solution a design parameter. Such a parameter is then traded off with the number of scenarios, which is clearly linked to computational complexity. The following assumption is required.
\begin{assumption}
The $N_{\delta}$ scenarios are independent and identically distributed (i.i.d.)
	\label{eq:assumptions}
\end{assumption}
Under Assumption \ref{eq:assumptions}, the following theorem holds (\cite{CAMPI2009149}).
\begin{thm}\label{Theorem}
	Consider a ``violation parameter'' $\epsilon \in (0,1)$ and a ``confidence parameter'' $\beta \in (0,1)$. If
	\begin{equation}
	N_{\delta} \geq \frac{2}{\epsilon}\left(\ln\left(\frac{1}{\beta}\right) + d -1 \right)
	\label{eq:scenarios_required}
	\end{equation}
	where $d$ is the number of optimization variables in \eqref{eq:scenario_samples}, then, with probability no smaller than $1-\beta$, $h^*$ satisfies all constraints in $\Delta$ with probability $1-\epsilon$, i.e.
	\begin{equation}
	Pr(f(\delta) > h^* ) \leq \epsilon
	\label{eq:scenario_probability}
	\end{equation}
\end{thm}
Notably, the confidence parameter $\beta$ can be very close to zero without scaling the required scenarios dramatically, see \eqref{eq:scenarios_required}, so that \eqref{eq:scenario_probability} is guaranteed with probability arbitrarily close to 1 (e.g., $1-10^{-7}$). In summary, the scenario approach allows one to certify the probability of violation of a found solution against new and unseen realizations of uncertainty, regardless of the probability distribution as long as it is the same one and samples are taken independently. Note that the bound \eqref{eq:scenarios_required} has the advantage of being explicit, however it is not tight: a tight and less conservative (but implicit) bound is also available from the theory (\cite{campi2018introduction}) and can be computed easily by numerical inversion.\\
We show next how this theory can be exploited in the training of ESN.
\section{ESN training with optimality guarantees} \label{opt}
%
%
When training  an ESN, the function $f(\delta)$ corresponds to the FIT performance obtained after training a network whose randomly chosen parameters (i.e. $W_x$, $W_u$, and $W_y$) correspond to the uncertain variables $\delta$.
This means that  the corresponding optimization problem \eqref{eq:scenario_samples} features just one optimization variable, i.e. $h$, hence $d=1$ in \eqref{eq:scenarios_required}. 
In view of this consideration, we propose to modify the tuning method described in Section~\ref{Preliminaries} as follows:
\begin{enumerate}[(a)]
	\item collect from the plant the, possibly normalized, input and output sequences $u_{sys}$ and $y_{sys}$;
	\item define the system order $n+n_u$;
	\item select the parameters $\beta$ and $\epsilon$. Compute the number of corresponding required scenarios $N_{\delta}$ according to~\eqref{eq:scenarios_required} with $d=1$;
	\item for each scenario $j=1,\dots,N_{\delta}$ repeat points \ref{Step 3}-\ref{Step 8} of the tuning method described in Section \ref{Preliminaries}, ending with $W_{out,j}$;
	\item using the available validation dataset, compute $FIT_j$ according to equation \eqref{FIT};
	\item take $j^*=\arg\max\limits_{j=1,\dots,N_{\delta}}FIT_j$;
	\item select $W_{x,j^*}$, $W_{u,j^*}$, $W_{y,j^*}$, $W_{out_1,j^*}$, and $W_{out_2,j^*}$ as the optimal parameters for model \eqref{eq:RNN}.
\end{enumerate}
%
%
%

Using the scenario optimization theory, we  directly obtain sound optimality guarantees on the model selected according to the procedure described above. In fact, in view of Theorem~\ref{Theorem}, it is possible to conclude that, if we  train a new model 
with random parameters $W_x$, $W_u$, $W_y$, the probability that the latter outperforms the one obtained with our procedure is smaller than $\epsilon$  (with confidence $(1-\beta) \approx 1$). To give a practical example, with a confidence parameter $\beta=10^{-7}$ and  violation parameter $\epsilon= 0.05= 5\%$, the resulting number of training instances that needs to be carried out  is $N_{\delta}=645$, which is very manageable. Most importantly, note that such a number of samples \emph{is independent from the complexity of the network}: in the example, we will need these 645 training instances no matter if the model order of the considered ESN is 10, 100, or 10$^6$.

So far, the approach has been presented by considering maximization of the FIT. This is without loss of generality: the procedure can be applied to any other objective function and either to minimization or maximization. For example, one can set up the equivalent procedure by considering the RMSE, in which case the problem is a minimization one, and one shall pick the ESN that scores the smallest RMSE among the $N_{\delta}$ trained instances.

Finally, note that the same reasoning also applies when one wants to derive guarantees on the worst-performing network, e.g. to compute a probabilistic lower bound to the FIT value or upper bound to the RMSE. In fact, it is enough to pick the worst-performing ESN among the trained ones, instead of the best-performing one, and the same probabilistic guarantees are obtained (i.e., with high probability, a newly trained network will score no worse than the selected one with probability $1-\varepsilon$).
\begin{remark}
	Notably, the proposed approach can be easily extended to other neural network structures, whenever the training procedure entails a random part. 
\end{remark}

\section{Simulation results}
In this section the described procedure is applied to assess the quality of ESN trained models of a simulated plant. Specifically, the non-linear benchmark employed in our simulations corresponds to the \emph{pH} neutralization process, typically exploited for the purification of waste waters. The aim of the process is to maintain the $pH$ of the solution in the principal tank at a neutral level, i.e. $pH=7$.\newline
\begin{figure}[h]
	\centering
	\includegraphics[scale=0.5]{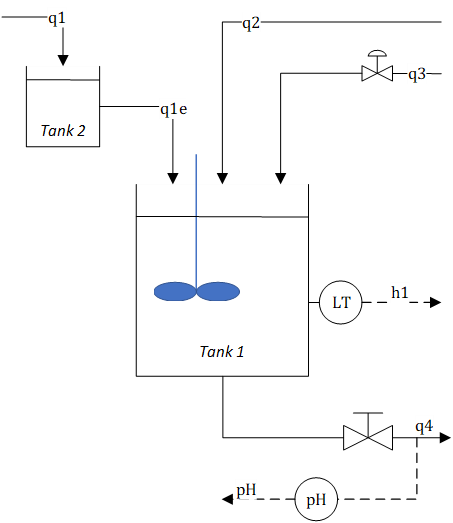}
	\caption{ $pH$ neutralization system scheme}
	\label{fig: sistemaph}
\end{figure}
In Figure \ref{fig: sistemaph} a simplified scheme of the process is reported; as it is possible to see, the system is composed of two tanks: a principal one, also known as reactor tank, in which the main transformation occurs, and an acid tank, which is fed by an acid stream $q_1$.
The reactor has three input flows, namely the acid stream $q_{1e}$ (output of the acid tank), the buffer flow $q_2$ (conjugate of acid-base pair solution) and the alkaline stream $q_3$, and it has one output flow $q_4$, which is the final solution where the $pH$ is measured and controlled.
Furthermore, the system is endowed with: a level sensor of the liquid inside the reactor, a sensor for the measurement of output solution $pH$ and an agitator that allows to keep the fluid characteristics constant in the liquid volume.\\
It is possible to derive a simplified state-space model of the process thanks to the following physical considerations:
\begin{itemize}
	\item the dynamics of the acid tank are much faster than those of the reactor tank, hence the input flow $q_{1e}$ is considered to be equal to $q_1$;
	\item the input flow-rate $q_{1e}$ is considered constant;
	\item the input flow-rate $q_2$ is considered as unmeasured disturbance.
\end{itemize}
The resulting physical model of the process has one control input regulated by a valve $u = q_3$, one output $y=q_4$, one disturbance $d=q_2$ and three states, that are two ions concentrations and the solution height inside the reactor, $x= [W_{a4}\; W_{b4}\; h_1]^T$. Ultimately the process is described by the following differential state-space equations with a constraint, as reported in \cite{4790490}:
\begin{equation}
\begin{aligned}
\dot{x}(t)= f_1(x(t))+f_2(x(t)) & u(t)+f_3(x(t))d(t)	\\[1.5ex]
c(x(t),y(t)) & = 0
\end{aligned}
\label{eq:processEq1}
\end{equation}
where
\begin{equation}
\scalebox{0.8}{$
	\begin{aligned}
	f_1(x(t)) & = \left[\frac{q_{1}}{A_{1}x_{3}}(W_{a1}-x_{1}),\frac{q_{1}}{A_{1}x_{3}}(W_{b1}-x_{2}),\frac{1}{A_{1}}(q_{1}-C_{v4}(x_{3}+z)^{n})\right]^{T}\\[1.5ex]
	f_2(x(t)) & = \left[\frac{1}{A_{1}x_{3}}(W_{a3}-x_{1}),\frac{1}{A_{1}x_{3}}(W_{b3}-x_{2}),\frac{1}{A_{1}}\right]^{T} \\[1.5ex]
	f_3(x(t)) & = \left[\frac{1}{A_{1}x_{3}}(W_{a2}-x_{1}),\frac{1}{A_{1}x_{3}}(W_{b2}-x_{2}),\frac{1}{A_{1}}\right]^{T}
	\notag
	\end{aligned}$}
\label{eq:processEq2}
\end{equation}
and
\[
\begin{array}{ll}
c(x,y) & = x_{1}+10^{y-14}+10^{-y}+x_{2}\frac{1+2\cdot 10^{y-pK_{2}}}{1+10^{pK_{1}-y}+10^{y-pK_{2}}}
\end{array}
\]
and $pK_{i}$ is the $i^{th}$ dissociation constant of the weak acid $H_{2}CO_{3}$.
In Table \ref{tab:Tab3} we report the nominal values of the model parameters, where $[M]=[\frac{mol}{L}]$.
\begin{table}[h]
	\centering	
	\caption{Nominal operating conditions of the $pH$ system}
	\begin{tabular}{lll}
		\toprule
		$z = 11.5\,cm$ 			& $W_{a1} = 3.00\cdot10^{-3}\,M$ &$q_{1} = 16.6\,mL/s$  \\
		$C_{v4} = 4.59$ 	& $W_{b1} = 0.00\,M$ &$q_{2} = 0.55\,mL/s$  \\
		$n = 0.607$ 			& $W_{a2} = -0.03\,M$ &	$q_{3} = 15.6\,mL/s$ \\
		$pK_{1} = 6.35$ 		& $W_{b2} = 0.03\,M$ &$q_{4} = 32.8\,mL/s$ \\
		$pK_{2} = 10.25$ 		& $W_{a3} = 3.05\cdot10^{-3}\,M$ &$A_{1} = 207\,cm^{2}$  \\
		$h_{1} = 14\,cm$		& $W_{b3} = 5.00\cdot10^{-5}\,M$ & $W_{a4} = -4.32\cdot10^{-4}\,M$\\
		$pH = 7.0$& $W_{b4} = 5.28\cdot10^{-4}\,M$\\
		\bottomrule
	\end{tabular}
	\label{tab:Tab3}
\end{table}

\subsection{Optimality guarantees \& validation}
\begin{figure}[h]
	\centering
	\includegraphics[scale=0.25]{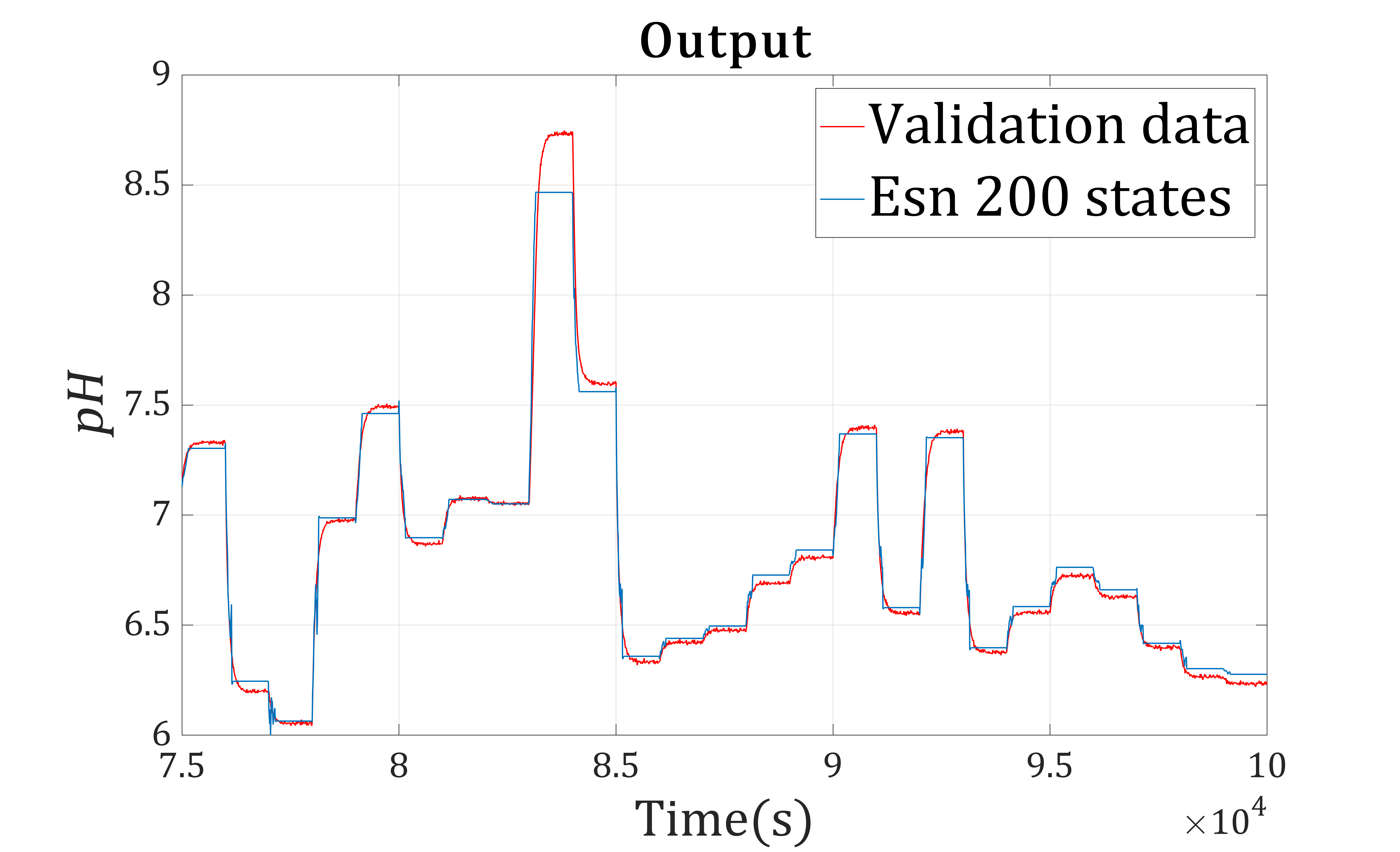}
	\caption{Validation dataset: sampled output of the real plant (red line) vs. ESN predictions with 200 states (blue line)}
	\label{F: output}
\end{figure}

In this section we present the simulation results produced by the application of the proposed procedure.\newline
The optimality guarantees of this approach are derived and tested by sampling a suitable number of instances $N_\delta$. The design parameters are given in Table \ref{T: parameters}.
\begin{table}[h]
	\centering	
	\caption{Design parameters for the scenario approach}
	\renewcommand\arraystretch{1.1}
	\begin{tabular}{cll}
		\toprule
		\multicolumn{1}{l}{Parameter}	& \multicolumn{1}{c}{Description}	& Value		\\
		\midrule
		$\beta$							& confidence parameter				& $10^{-7}$	\\
		$\epsilon$ 						& violation parameter				& $0.05$	\\
		$d$ 							& optimization variables			& $1$		\\
		$N_\delta$ 						& number of instances				& $645$		\\
		\bottomrule
	\end{tabular}
	\label{T: parameters}
\end{table}
First of all, we have employed a simulator of the real plant, including an output-measurement disturbance to retrieve realistic data useful for training the ESN models. A Multilevel Pseudo-Random Signal (MPRS), simulating the input flow rate $q_3$, has been used to excite the plant over the whole operating conditions using a switching period of $1000$ seconds, which is longer than the real system settling time, and an amplitude included in the interval $[12.7 , 16.7]$ mL/s. The corresponding output of the process takes values in between 6 and 8.65, see the validation data in Figure \ref{F: output}.

The sampling time for input and output signals used to train the different ESN of the scenario approach is $T_s=10$ s to empirically obtain 30 samples in the settling time of the system's step response.

\begin{figure}[h]
	\centering
	\subfigure[][Scenario sampling: examples of computation of the best case FIT (top figure), and worst case RMSE (bottom figure).\label{F: Sampling}] {\includegraphics[scale=0.32]{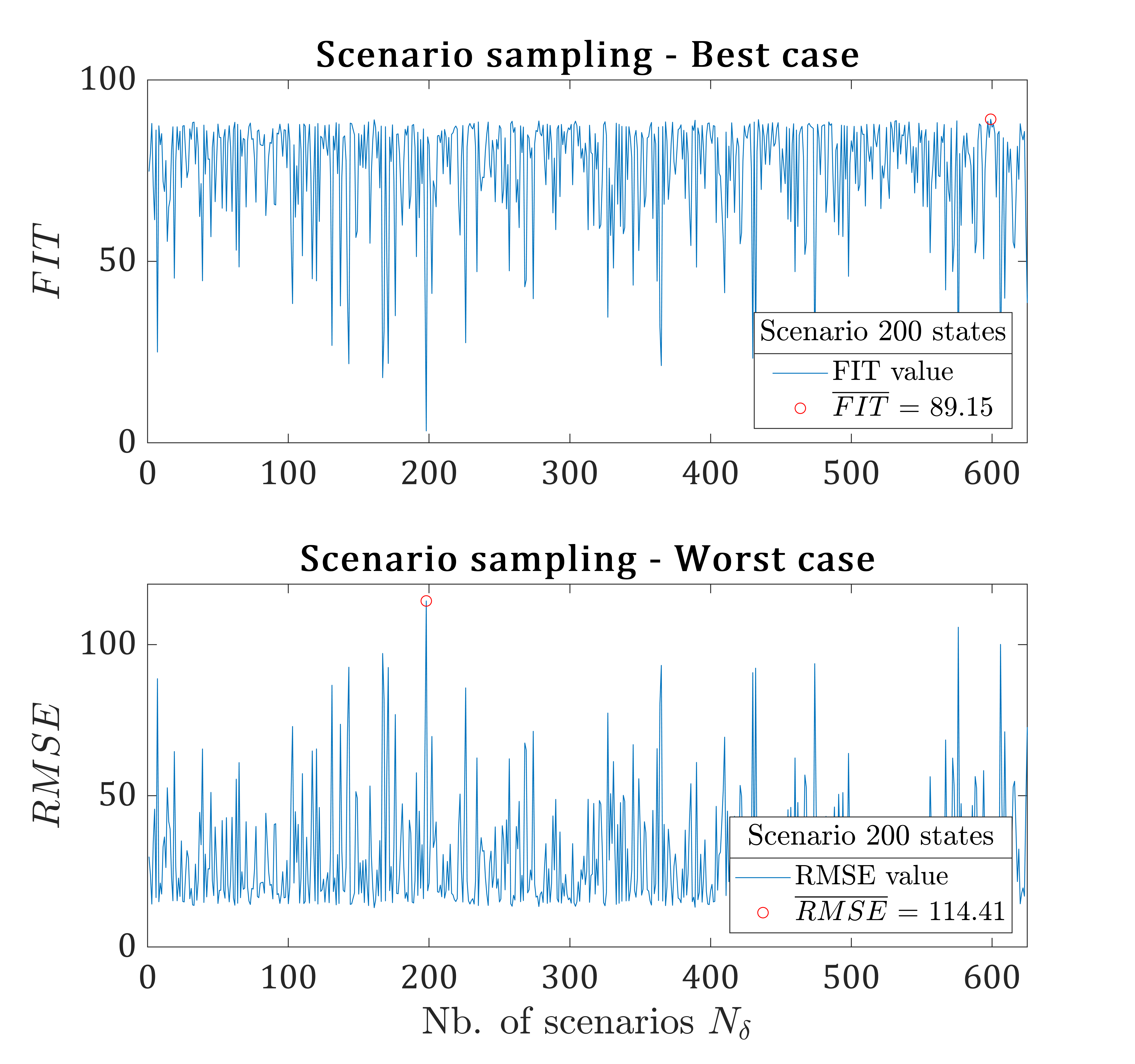}}\\
	\subfigure[][Scenario testing: example of empirical test of the best case FIT (top figure), and  worst case RMSE (bottom figure).\label{F: Testing}] {\includegraphics[scale=0.32]{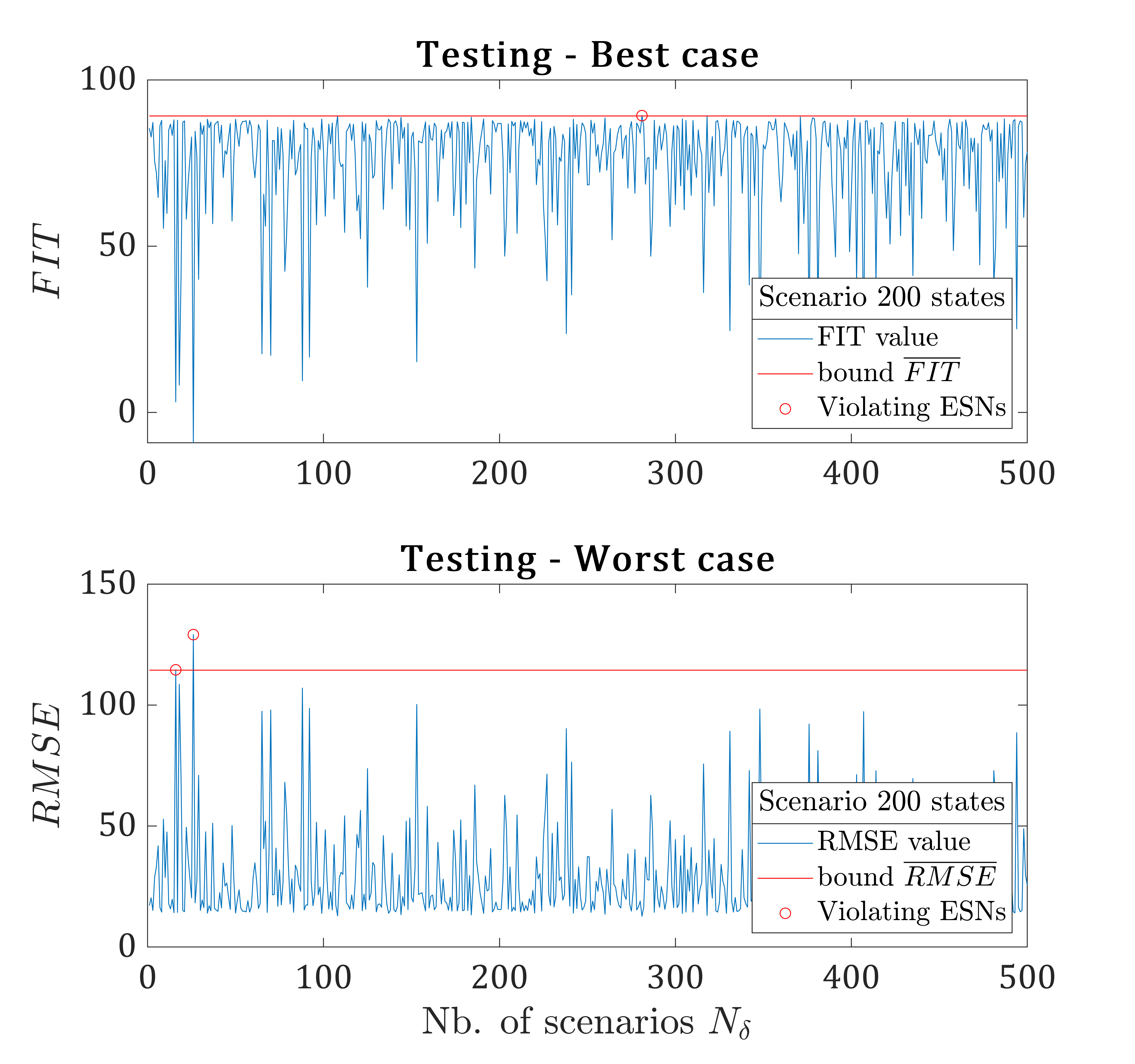}}
	\caption{Scenario approach: example of sampling and testing of an ESN of order 200.}
	\label{F: TestScen}
\end{figure}

Subsequently, according to the algorithm proposed in Section~\ref{opt}, we trained $N_\delta = 645$ ESN models computing all the respective values of $RMSE_j$ and $FIT_j$, $j=1,\dots,645$ on validation data, for different values of the model order (from 10 to 500 states). We have finally computed the values $\overline{FIT}$ and $\overline{RMSE}$, which correspond to the best FIT and the worst RMSE value, respectively, over the 645 instances. \newline
Then, to verify empirically the validity of the guarantees provided by the theory, we collected other 500 new scenarios, trained as many ESNs and evaluated the associated RMSE and FIT values over the same validation dataset. In this way, we could empirically test the conservativeness of the bounds $\overline{RMSE}$ and $\overline{FIT}$ and check their consistency with the probability $\epsilon$. In other words, we have verified that no more than $\epsilon$-cases over the total amount of new scenarios score a RMSE lower than $\overline{RMSE}$ or a FIT higher than $\overline{FIT}$.\\
In Table \ref{T: Testing} we present the absolute number of violations over the total amount of new instances considered in the testing phase (500) for different model orders. In all the cases the probabilistic bounds are verified, as the maximum number of ESNs that violate the constraints are respectively $3/500=0.6\%$ for the FIT and $4/500=0.8\%$ for the RMSE. Note that these values  are definitely much smaller than the guaranteed violation probability $\epsilon=5\%$. A possible reason for this is that, for a given sample of the internal weights, the training procedure still tries to optimize the predictive capability of the network with the least squares estimation of the output layer.\newline
Figure \ref{F: TestScen} shows the results obtained with the scenario approach, both for the best and worst cases, using ESNs characterized by 200 internal units. In particular, Figure \ref{F: Sampling} reports the sampling procedure performed to obtain the best $\overline{FIT}$ and the worst  $\overline{RMSE}$, while Figure \ref{F: Testing} displays the testing phase on new collected instances, where the thresholds (red continuous lines) indicate the optimal values derived in the sampling phase, and the circles indicate for the ESNs that violate the bounds enforced by such optima.

\begin{table}[h]
	\centering
	\caption{Testing violations over 500 new instances}
	\renewcommand\arraystretch{1.1}
	\begin{tabular}{clll}
		\toprule
		\multicolumn{1}{l}{Nb. of ESN states}	& \multicolumn{1}{c}{Best case}	& \multicolumn{1}{c}{Worst case}	& \multicolumn{1}{c}{$\overline{FIT}$}		\\
		\midrule
		10		& 0/500 = 0.0\%		& 0/500 = 0.0\%	& 82.30\\
		20		& 0/500 = 0.0\%		& 0/500 = 0.0\%	& 84.57\\
		30		& 1/500 = 0.2\%		& 0/500 = 0.0\%	& 85.40\\
		40		& 1/500 = 0.2\%		& 1/500 = 0.2\%	& 86.10\\
		50		& 0/500 = 0.0\%		& 0/500 = 0.0\%	& 87.08\\
		60		& 3/500 = 0.6\%		& 4/500 = 0.8\%	& 87.10\\
		70		& 0/500 = 0.0\%		& 1/500 = 0.2\%	& 87.73\\
		80		& 0/500 = 0.0\%		& 1/500 = 0.2\%	& 88.06\\
		90		& 3/500 = 0.6\%		& 0/500 = 0.0\%	& 87.79\\
		100		& 0/500 = 0.0\%		& 0/500 = 0.0\%	& 87.95\\
		120		& 0/500 = 0.0\%		& 3/500 = 0.6\%	& 88.42\\
		140		& 0/500 = 0.0\%		& 4/500 = 0.8\%	& 88.75\\
		160		& 0/500 = 0.0\%		& 0/500 = 0.0\%	& 89.04\\
		180		& 0/500 = 0.0\%		& 4/500 = 0.8\%	& 89.20\\
		200		& 1/500 = 0.2\%		& 2/500 = 0.4\%	& 89.15\\
		250		& 0/500 = 0.0\%		& 0/500 = 0.0\%	& 89.88\\
		300		& 0/500 = 0.0\%		& 3/500 = 0.6\%	& 90.15\\
		350		& 0/500 = 0.0\%		& 2/500 = 0.4\%	& 90.16\\
		400		& 0/500 = 0.0\%		& 1/500 = 0.2\%	& 90.11\\
		450		& 3/500 = 0.6\%		& 0/500 = 0.0\%	& 89.85\\
		500		& 0/500 = 0.0\%		& 1/500 = 0.2\%	& 90.08\\
		\bottomrule
	\end{tabular}
	\label{T: Testing}
\end{table}

\section{Conclusions}
We presented an application of the scenario approach to train Echo State Networks, a popular class of recurrent neural networks. 
First, the established training algorithm used to derive a state-space model of ESN is reported, where the parameters of the state equations are randomly generated. Then, this algorithm is modified to obtain the optimal network - with sound guarantees - according to the scenario approach. Eventually, the approach has been tested in order to empirically evaluate the guarantees on the FIT and RMSE values computed on a validation dataset for the models identified with ESNs. A novel set of scenarios confirmed the reliability of the solutions derived through the application of the proposed algorithm.\newline
Future work is concerned to the extension to the case of constraints removal and the adaptation of neural networks with online data.

\bibliography{Bibliografia}             
\bibliographystyle{plain}

\end{document}